\documentstyle[12pt,epsfig]{article}
\newcommand{\beq}{\begin{equation}}
\newcommand{\eeq}{\end{equation}}
\newcommand{\ba}{\begin{array}}
\newcommand{\ea}{\end{array}}
\newcommand{\beqa}{\begin{eqnarray}}
\newcommand{\eeqa}{\end{eqnarray}}
\newcommand{\lsim}{\stackrel{<}{_\sim}}
\newcommand{\gsim}{\stackrel{>}{_\sim}}
\newcommand{\PL}[3]{{ Phys. Lett.}       {\bf #1} {(19#2)} {#3}}

\newcommand{\PRL}[3]{{ Phys. Rev. Lett.} {\bf #1} {(19#2)} {#3}}
\newcommand{\PR}[3]{{ Phys. Rev.}        {\bf #1} {(19#2)} {#3}}
\newcommand{\NP}[3]{{ Nucl. Phys.}       {\bf #1} {(19#2)} {#3}}
\newcommand{\ZP}[3]{{ Z.  Phys.}         {\bf #1} {(19#2)} {#3}}
\newcommand{\no}{\noindent}

\begin{document}
\begin{titlepage}
\begin{flushright}
LNF-96/075(P) \\
hep-ph/9612429 \\
December 1996\\
\end{flushright}
\begin{center}
 
\vglue 3.0 true cm
 
{\large \bf A status report concerning theoretical \\
            predictions for various kaon decays*}
\vspace*{1 cm}
\author{Gino Isidori\\
INFN, Laboratori Nazionali di Frascati, 
P.O. Box 13, I-00044 Frascati, Italy } 

\vspace*{1cm}
{\bf Gino Isidori }

\vspace{.5cm}
INFN, Laboratori Nazionali di Frascati \\ 
P.O. Box 13, I--00044 Frascati, Italy

\vfill
{\bf Abstract} \\
\end{center}
\no
A short overview of theoretical predictions for various
kaon decays is presented.  Par\-ti\-cu\-lar attention is devoted to 
pure and radiative nonleptonic
decays in the framework of  Chiral Perturbation Theory.
The relevance of KLOE's future results  \cite{Aloisio,Handbook} 
to improve our knowledge of kaon physics and more generally
of the Standard Model at low energy is also emphasized.

\vfill
\noindent * Invited Talk presented at DAPHCE96, Frascati, Italy,
11 -- 14 November 1996, to appear in the proceedings.

\end{titlepage}

\section{Introduction}
\no
Kaon decays offer a unique possibility to test strong, 
weak and electromagnetic interactions (i.e. the Standard Model)
at low energies.  
At the same time, through flavour--changing neutral current 
and $CP$--violating processes, kaon decays are sensitive to  
new physics up to the TeV scale.  The interest and the 
variety of such decays (see e.g. Tab.~\ref{tab:1}) is definitely too 
large to be covered in this talk and we refer to some recent 
reviews \cite{Handbook,Ecker,Pich,deRafael,Noi} for a comprehensive analysis.

\begin{table}[h]
\newlength{\digitwidth} \settowidth{\digitwidth}{\rm 0}
\catcode`?=\active \def?{\kern\digitwidth}
\caption{Incomplete list of what we can learn from kaon decays.
 ${\cal L}_S$  and  ${\cal L}_W^{|\Delta S|=1}$ denote strong
 and weak nonleptonic chiral Lagrangians, respectively. }
\label{tab:1}
\begin{tabular*}{\textwidth}{@{}l@{\extracolsep{\fill}}ccc}\hline
channel  &  ${\cal L}_S$  & 
 ${\cal L}_W^{|\Delta S|=1}$ & other (CP, CPT, $U_{CKM}$) \\ \hline\hline
 $l\nu\gamma$  & determination of              &  &  bounds on         \\ 
 $l\nu e^+e^-$ & $L_{9}$ + $L_{10}$  &  &  tensor couplings  \\ \hline
 $\pi l\nu$     & determination of   &  &  bounds on t. c., \\ 
               &  $\lambda_+$ and $\lambda_0$   &  & T$\!\!\!\! /$~~,  
 CPT tests \\ \hline
 $\pi l\nu\gamma$  & test of the  &  &      \\ 
                   & WZW sector   &  &      \\ \hline
 $\pi\pi l\nu\gamma$  & $<0|\bar{q}q|0>$      &  &      \\ 
                     & ($\pi\pi$ phase shifts)  &  &      \\ \hline\hline
 $2\pi$  &  & $O(p^4)$ operators  & $\epsilon$, $\epsilon'/\epsilon$,  
            CPT tests               \\ \hline
 $3\pi$  & $3\pi$ phase shifts  & $O(p^4)$ operators  & $\delta g/g$ , 
         $\epsilon'_{000}$, $\epsilon'_{+-0}$\\
 & ($K_{L,S}\to \pi^+\pi^-\pi^0$) &    
   & \\ \hline\hline
 $\gamma\gamma$  &   & $O(p^6)$ operators  & $\epsilon'_\perp$, 
        $\epsilon'_\parallel$   \\ \hline
 $\pi\gamma\gamma$  &   & $O(p^4)$ +  $O(p^6)$ operators,  
                  & intermediate role in \\ 
                  &  & unitarity corrections  & $K\to\pi e^+e^-$ \\ \hline
 $2\pi\gamma$   &   & $O(p^4)$ operators,   &   
                $\epsilon'_{+-\gamma}$,   $\delta \Gamma/\Gamma$  \\
 $3\pi\gamma$   &   &weak anomalous sector&  \\ \hline\hline
 $\mu^+\mu^-$  &   & $O(p^6)$ operators  &  $|V_{td}V_{ts}^*|$ from \\ 
 $e^+ e^- \gamma(\gamma^*)$ &  &  & $K_L \to \mu^+\mu^-|_{SD}$ \\ \hline
 $\pi e^+e^-$  &   & $O(p^4)$ operators  &  CP$\!\!\!\!\!\! /$~~-dir. in  \\ 
               &   &   & $K_L \to \pi^0 e^+e^-$ \\ \hline\hline
 $\pi\nu\bar{\nu}$  &   &   &  $|V_{td}V_{ts}^*|$  
($K^+ \to \pi^+ \nu\bar{\nu} $)  \\  &   &   & CP$\!\!\!\!\!\! /$~~-dir.
($K_L \to \pi^0 \nu\bar{\nu} $)  \\ \hline\hline
\end{tabular*}
\end{table} 

The natural tool to analyze the Standard Model at low
energies is Chiral Perturbation Theory \cite{Leutwyler} (CHPT), 
in its $SU(3)$ version if we are interested in processes 
involving the strange quark. Within this 
framework kaon decays play a twofold role.
On one side semileptonic transitions let us to explore the strong 
sector of the chiral Lagrangian, answering to fundamental questions 
like the chiral behaviour of the quark  condensate. 
On the other side nonleptonic and radiative
decays help us to understand the chiral realization of the weak
four--quark effective hamiltonian, unravelling possibly 
short--distance effects.

In this talk we concentrate on nonleptonic
processes. Few interesting top\-ics in se\-mi\-lep\-to\-nic decays 
are just mentioned whereas 
the problem of $CP$ violation in $K\to 2 \pi$
is completely omitted. Both these
subjects are discussed  elsewhere in these
proceedings \cite{Leutwyler,Pennington,Maiani}.

\section{Semileptonic decays}
\no
The $SU(3)$ version of CHPT is based on the assumption that the 
$SU(3)_L\times SU(3)_R$ symmetry of the QCD Lagrangian in the 
chiral limit ($m_u=m_d=m_s$) is spontaneously broken and 
that the corresponding Goldstone modes can be identified with
the octet of light pseudoscalar mesons 
($\pi$, $K$, $\eta_8$). The approach is then to build 
the most general Lagrangian consistent with the 
$SU(3)_L\times SU(3)_R$ symmetry in terms of the Goldstone boson 
fields, and add to it the soft--breaking terms induced by quark masses.
Such a Lagrangian is not renormalizable and in principle contains 
an infinite number of arbitrary constants. Nevertheless, if we are
interested in low energy processes we can expand
the transition amplitudes up to a given order in powers of pseudoscalar
masses and momenta and consider only a finite number of 
such constants. 

\begin{table}[h]
\caption{Occurrence of the low--energy coupling constants 
$L_1,\ldots,L_{10}$ and of
the anomaly in kaon semileptonic decays \protect\cite{semi}. 
In $K_{\mu 4}$ decays,
the same constants as in the electron mode (displayed here) occur. In
addition, $L_6$ and $L_8$ enter in the channels
$K^+\rightarrow \pi^+\pi^-\mu^+\nu_\mu$ and
$K^+\rightarrow \pi^0\pi^0\mu^+\nu_\mu.$ }
\vskip 0.1 in
\label{tab:2}
\begin{tabular*}{\textwidth}{@{}c@{\extracolsep{\fill}}ccccccc}\hline
 & & & & & $K^+\rightarrow$ & $K^+\rightarrow$ & $K^0\rightarrow$
\\     &
$K_{l2\gamma}$ &
$K_{l2ll}$    &
$K_{l3}$    &
$K_{l3\gamma}$&
$\pi^+\pi^-e^+\nu_e$&
$\pi^0\pi^0e^+\nu_e$&
$\pi^0\pi^-e^+\nu_e$ \\ \hline
 $L_1$ &      &      &      &      & $\times$ &$\times$ &
\\
 $L_2$&      &      &      &      & $\times$ &$\times$ &
\\
 $L_3$ &      &      &      &      & $\times$ &$\times$ & $\times$
\\
 $L_4$ &      &      &      &      & $\times$ &$\times$ &
\\
 $L_5$&       &      &      &      & $\times$ &$\times$ &$\times$
\\
 $L_9$&      &$\times$&$\times$&$\times$& $\times$ &$\times$ &$\times$
\\
 $L_9+L_{10}$ &$\times$&$\times$&      &$\times$&       & &
\\  \hline
Anomaly&$\times$&$\times$& &$\times$&$\times$&&$\times$ \\
\hline
\end{tabular*}
\end{table}

Gauging the global symmetry $SU(3)_L\times SU(3)_R$ leads to 
describe, in terms of the `strong'  
chiral Lagrangian, also semileptonic transitions at $O(G_F)$. 
The relevance of these processes for the determination of
the constants $L_i$ appearing in the $O(p^4)$ 
(next--to--leading order) chiral Lagrangian \cite{GL} 
is shown in Tab.~\ref{tab:2} \cite{semi}.
Since to date all the $L_i$ have been determined phenomenologically,
the picture of semileptonic decays is complete and 
offers the possibility of precise and interesting 
tests of the Standard Model. 

\begin{figure}[t]
\begin{minipage}[t]{77mm}
       \begin{center}
        \setlength{\unitlength}{1.truecm}
       \begin{picture}(3.0,6.5)
       \put(-2.3,-0.5){\includegraphics{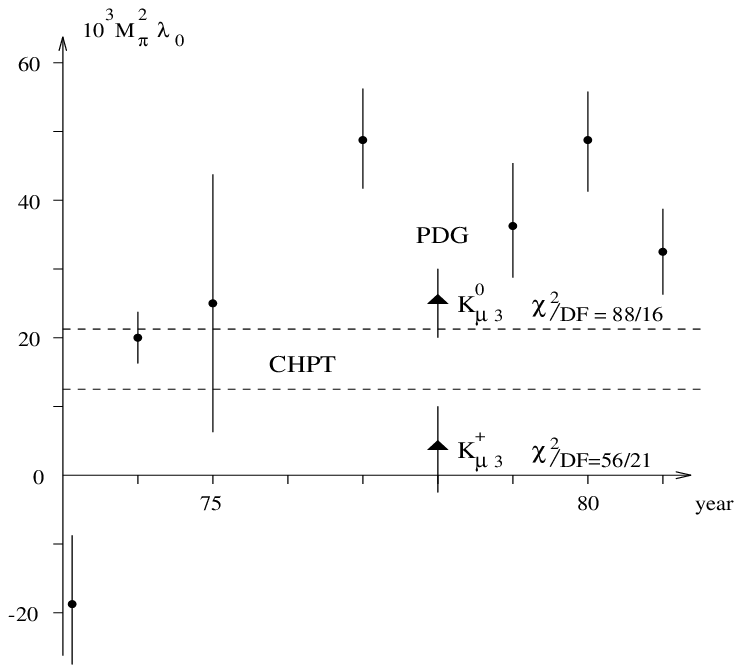}}
       \end{picture}
       \end{center}
    \caption{Comparison between the experimental data and the 
CHPT prediction for the scalar form factor $\lambda_0$ 
\protect\cite{semi2,Gasserl0}. Dots denote 
$K^0_{\mu 3}$ measurements; the two triangles indicate the PDG averages 
 for $K^0_{\mu 3}$ and $K^+_{\mu 3}$ \protect\cite{PDG}; 
the two dashed lines show the CHPT prediction $\lambda_0=0.17 \pm 0.04$.}
\label{fig:l0}
\end{minipage}
\hspace{\fill}
\begin{minipage}[t]{77mm}
      \begin{center}
       \setlength{\unitlength}{1truecm}
       \begin{picture}(3.0,6.5)
       \put(-2.8,-0.5){\includegraphics{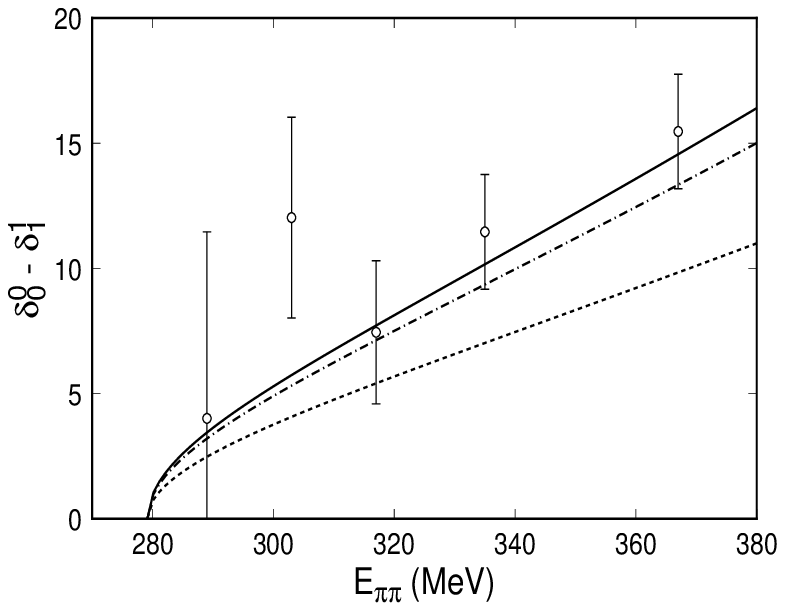}}
       \end{picture}
       \end{center}
    \caption{The phase shift 
difference $\delta_0^0-\delta_1^1$ (in degrees)
as a function of the $\pi\pi$ 
center--of--mass energy. 
Dotted, dash--dotted and full lines denote respectively 
tree, one--loop and two--loop  CHPT results~\protect\cite{Bijnens}. 
The bars indicate the results of the $K_{e4}$ experiment of 
Rosselet et al. \protect\cite{Ross}.}
\label{fig:pp}
\end{minipage}
\end{figure}

Among the various tests it is worthwhile to
mention at least two examples of particular interest for KLOE: 
the determination of the scalar form--factor 
$\lambda_0$ in $K_{\mu3}$ and the determination of $\pi\pi$ phase shifts near 
threshold in $K_{e4}$ (Figs.~\ref{fig:l0}--\ref{fig:pp}). In both cases there 
are firm CHPT predictions 
whereas the present experimental situation is not clear \cite{semi2}. In the 
case of $\pi\pi$ phase shifts, recently calculated up to two loops, i.e. at 
$O(p^6)$, both in CHPT \cite{Bijnens} and in `generalized CHPT' 
\cite{Stern}, accurate data could tell us which is the behaviour
of the quark condensate in the 
chiral limit \cite{Leutwyler,Pennington}. 

\section{Nonleptonic decays}

\no
The operator product expansion (OPE)
let us to compute  nonleptonic $|\Delta S|=1$ transitions   
at low energies ($\mu \ll M_W$)  by means of an effective four--fermion
Lagrangian \cite{Buras,Ciuchini}: 
\beqa
{\cal L}_{eff}^{|\Delta S|=1} &=&
 {4 G_F \over  \sqrt{2}} V_{ud}{V_{us}}^*
 \sum_{i} C_i(\mu) O_i  + \mbox{\rm h.c.}. 
\label{ciuham}
\eeqa
The four--fermion operators $O_i$ contain only   
light fermion fields $\psi_l$ ($m_f < \mu$),
whereas the Wilson coefficients  $C_i(\mu)$ 
depend on the heavy degrees of freedom integrated out.
The renormalization scale $\mu$ introduces an artificial 
distinction between short-- and long--distance effects:
the $\mu$--dependence of the $C_i(\mu)$ that parametrizes
short-distance effects is cancelled by the 
$\mu$--dependence of the four--fermion  operator matrix elements 
between initial and final hadronic states.

The Wilson coefficients can be calculated 
reliably using renormalization group techniques down to 
$\mu \gsim m_c$,  where QCD is still in  a  perturbative
regime. Recently the $C_i(\mu)$ have been calculated at the
next--to--leading order \cite{Buras,Ciuchini} 
drastically reducing the theoretical  uncertainties 
on the short--distance part of Eq.~(\ref{ciuham}). The main source of
uncertainties in nonleptonic kaon decays is therefore the  
evaluation of the four-fermion operator matrix elements. 

There are two ways to approach this problem. On 
one side we can try to use non--perturbative techniques (lattice
QCD, $1/N_c$ expansion, resonance saturation, etc...) to estimate
the matrix elements of the operators $O_i$. On the other side, following
CHPT assumptions, we can use the symmetry 
properties of the four--fermion lagrangian under $SU(3)_L\times SU(3)_R$ 
to construct its realization in terms of the pseudo Goldstone--boson 
fields. The second solution is certainly less predictive, since a
number of low--energy constants must be introduced, but is 
the most systematic and in many cases the most reliable. 
Nonetheless the two approaches are complementary. Up to date  
non--perturbative techniques are not accurate enough to fix all the unknown 
couplings of the CHPT Lagrangian, but in future we may expect
that the two approaches will merge yielding a systematic and 
completely predictive picture of nonleptonic decays.

The lowest--order chiral realization of the four--fermion 
Lagrangian (\ref{ciuham}) is very simple
\beq
{\cal L}_W^{(2)} =  F^4 \left[  G_8  W_8^{(2)}
+  G_{27} W_{27}^{(2)} \right ] + \mbox{\rm h.c.}
\label{lagr2w}
\eeq
where $W_8^{(2)}$ and $W_{27}^{(2)}$ are  $O(p^2)$ operators
transforming under   
$SU(3)_L\times SU(3)_R$
as $(8_L,1_R)$ and $(27_L,1_R)$, respectively\footnote{~We neglect the 
suppressed $(8_L,8_R)$ electromagnetic--penguin operators (see 
Ref.~\protect\cite{Noi,Fabbrichesi} for a 
recent discussion about this point).}.
The two unknown couplings,
naively expected to be of the order of $\sin\theta_C \times G_F$, can 
be fixed from the $K\to 2\pi$ widths. In this case one finds
\beq
|G_8| \simeq 9.1 \times 10^{-6}\ \mbox{\rm GeV}^{-2}, \qquad\qquad
G_{27}/G_8  \simeq  1/18,
\eeq
where the suppression of the ratio $G_{27}/G_{8}$ 
is a consequence of the phenomenological enhancement 
of $\Delta I=1/2$ transitions.	

The Lagrangian (\ref{lagr2w}) let us to predict 
the decay amplitudes of $K\to 3\pi$, $K\to 2\pi\gamma$ and $K\to 3\pi\gamma$
in terms of $G_8$ and $G_{27}$. However, in many cases, the lowest order
predictions are not accurate enough to describe present data
and a complete $O(p^4)$ analysis is needed. This is even more evident for 
processes like  $K\to \pi l^+l^-$, $K\to \gamma\gamma$ and 
$K\to \pi\gamma\gamma$ where the lowest order predictions vanish.

\begin{figure}[t]
    \begin{center}
       \setlength{\unitlength}{1truecm}
       \begin{picture}(11.0,12.0)
       \put(-1.0,-2.0){\includegraphics{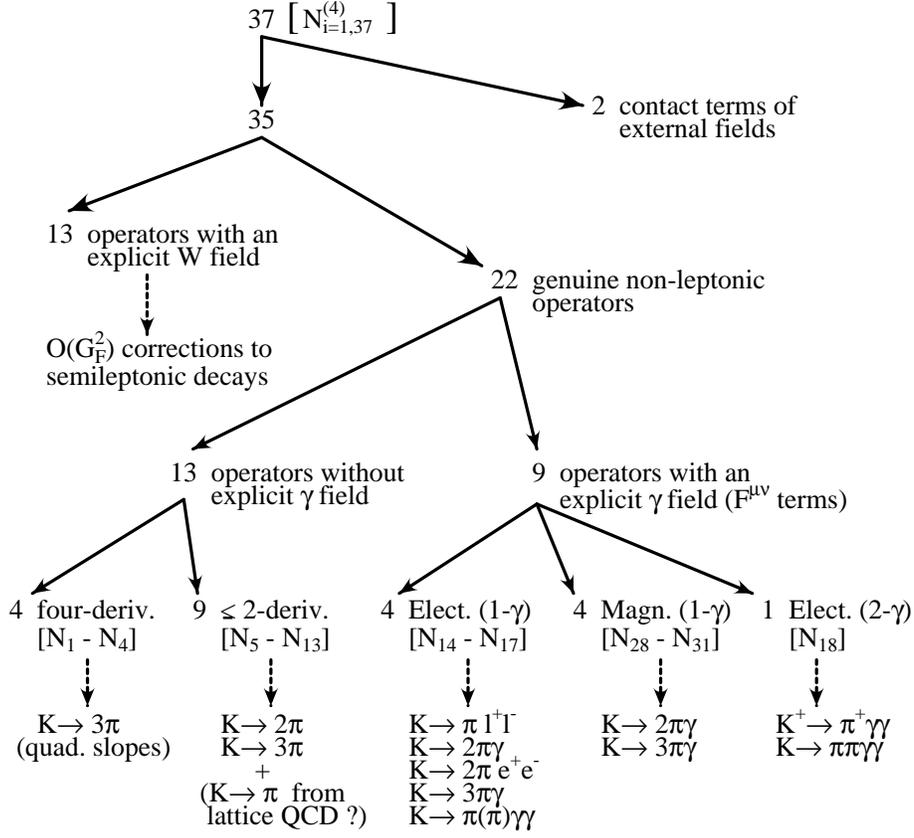}}
       \end{picture}
    \end{center}
    \caption{Role of the low--energy constants $N_i$ appearing in 
  the $O(p^4)$ nonleptonic octet Lagrangian of 
  Ref.~\protect\cite{EKW}.}
    \protect\label{fig:ct}
\end{figure}

The next--to-leading order 
structure of the nonleptonic Lagrangian is quite
complicated since the number of independent operators substantially 
increase. The analysis of such operators, started by
Ecker, Pich and de Rafael in the radiative sector \cite{EPR1,EPR3}, 
has been completed by Kambor, Missimer and Wyler
both for the octet and the 27-plet components \cite{KMWn}. 
In the case of the octet sector a useful basis has been introduced 
in Ref.~\cite{EKW}, where the number of independent operators has been
reduced to 37. As can be noticed from Fig.~\ref{fig:ct}, in the 
basis of Ecker, Kambor and Wyler \cite{EKW} the octet
operators are organized in such a way that only few of them 
occur in a definite nonleptonic process. In principle  
there are  enough  observables (particularly in the radiative sector)
both to disentangle several $N_i$ and to perform consistency  
checks of the CHPT approach \cite{Nellohand,Eckerrep}. 
Unfortunately present data do not allow
to fulfill this program but the situation will 
certainly change after the completion of the KLOE experiment. 
We stress that a better knowledge of the $N_i$ is not only
useful by itself but also to discriminate among the 
various non--perturbative hadronization models.

\subsection{$K \to 3 \pi$}

\begin{table}
\begin{minipage}[t]{73mm}
\caption{Number of independent isospin amp\-li\-tu\-des and 
$O(p^4)$ free parameters in $K\to3\pi$.}
\vskip 0.1 in
\label{tab:3}
\begin{tabular*}{\textwidth}{@{}c@{\extracolsep{\fill}}cccc}\hline
       & $\Delta I=1/2$ & $\Delta I=3/2$ & phases   \\ \hline \hline 
const. &    1           &     1          &   1      \\
       & ($\alpha_1$)   & ($\alpha_3$)   &         \\ \hline
linear &    1           &     2          &   3      \\
       & ($\beta_1$ )   & ($\beta_3,\gamma_3$ ) &   \\ \hline
quad.  &    2           &     3          &   -      \\
       & ($\zeta_1,\xi_1$) & ($\zeta_3,\xi_3,\xi'_3$) & \\ \hline\hline
$O(p^4)$ &  2   & 3 & - \\ free par. & & & \\ \hline
\end{tabular*}
\end{minipage}
\hspace{\fill}
\begin{minipage}[t]{82mm}
\caption{Experimental results and 
CHPT predictions for $K\to3\pi$ amplitudes.}
\vskip 0.1 in
\label{tab:4}
\begin{tabular*}{\textwidth}{@{}c@{\extracolsep{\fill}}cccc}\hline
$\quad\quad$ & 
$O(p^2)$ & $O(p^4)$ \protect\cite{Kamborslopes,KMWlett}
 & exp. fit \protect\cite{KMWlett} \\ \hline\hline
$\alpha_1$& $74.0$   & input    & $ 91.71\pm0.32$  \\ \hline
$\beta_1$ & $-16.5$  & input    & $-25.68\pm0.27$ \\ \hline
$\zeta_1$ & $-$      & $-0.47\pm 0.18$   & $-0.47\pm 0.15$ \\ \hline
$\xi_1$   & $-$      & $-1.58\pm 0.19$   & $-1.51\pm 0.30$ \\ \hline
$\alpha_3$& $-4.1$   & input    & $-7.36\pm 0.47$  \\ \hline
$\beta_3$ & $-1.0$   & input    & $-2.43\pm 0.41$ \\ \hline
$\gamma_3$& $1.8$    & input    & $2.26 \pm 0.23$  \\ \hline
$\zeta_3$ & $-$      & $-0.011\pm 0.006$  & $-0.21\pm 0.08$ \\ \hline
$\xi_3$  & $-$      &  $0.092\pm 0.030$  & $-0.12\pm 0.17$ \\ \hline
$\xi'_3$   & $-$      & $-0.033\pm 0.077$  & $-0.21\pm 0.51$ \\ \hline
\end{tabular*}
\end{minipage}
\end{table}

\no
$K \to 3 \pi$ amplitudes are usually
expanded in terms of the Dalitz Plot variables $X$ and $Y$
\cite{Nellohand}:
\beq
A(K \to 3 \pi) = a + b Y + b' X + O(X^2,XY,Y^2),
\label{slopes}
\eeq
where
\beq
X= { s_1 - s_2 \over M_\pi^2 }, \qquad  
Y= { s_3 - s_0 \over M_\pi^2 }, \qquad  s_i=(p_k-p_{\pi_i})^2, 
\qquad s_0=(s_1+s_2+s_3)/3~.
\eeq 
Present data are well described by an expansion up to 
quadratic terms (higher powers of 
$X$ and $Y$ belong to high angular momentum states). 
The $O(p^2)$ CHPT predictions are non--vanishing only for
constant and linear terms \cite{Cronin}. Quadratic slopes and 
re--scattering phases are generated at $O(p^4)$ \cite{KMWlett}.
The assumption of isospin symmetry and 
the inclusion of re--scattering 
phases up to linear terms 
provide us with a  parametrization
of the decay amplitudes of the five $CP$--conserving processes, 
\beq  
K^\pm \to \pi^\pm \pi^\pm \pi^\mp,~K^\pm \to \pi^0 \pi^0 \pi^\pm,~
K_L \to \pi^+\pi^-   \pi^0,~K_L \to \pi^0 \pi^0 \pi^0,~
K_S \to \pi^+ \pi^-   \pi^0,
\eeq
in terms of 13 real parameters \cite{KMWlett,DIPP}
(see Tab.~\ref{tab:3}).

As can be noticed from Tabs.~\ref{tab:3}--\ref{tab:4}, at $O(p^4)$
it is possible to predict unambiguously $K\to3 \pi$
quadratic slopes and re--scattering phases. The first ones 
have been measured
with reasonable accuracy only in the $\Delta I=1/2$ sector and there the 
agreement with CHPT is good. An improvement in the
experimental determination of $K\to3 \pi$  slopes, together with a
theoretical analysis of  isospin--breaking effects, 
is definitely needed to analyze the 
 $\Delta I=3/2$ sector and thus to
understand better the so--called 
`$\Delta I=1/2$ rule'.

Re--scattering phases cannot be  
extracted by a simple analysis of $K\to 3\pi$ widths. 
The best way to experimentally access to the $3\pi$ phases is through
time--interference measurements in the neutral channel $K_{L,S} \to 
\pi^+\pi^-\pi^0$ \cite{DamPav,DIPP}.  This method has recently let to 
observe the rare $K_{S} \to \pi^+\pi^-\pi^0$
decay \cite{E621_cons,CPLEAR_cons} (a pure $\Delta I=3/2$ transition).
Unfortunately present data are 
affected by large errors and do not to allow to test $O(p^4)$
effects both in the real and in the imaginary 
parts of the amplitudes. A significant improvement on this kind of
measurements is expected at KLOE.

Another interesting aspect of $K\to 3\pi$ decays are the 
direct $CP$--violating asymmetries. In this case the CHPT approach
does not allow to make firm predictions since we have not a good control
on the imaginary parts of the low--energy constants. Nevertheless,
using the available information from $K\to 2 \pi$ 
(experimental and lattice QCD results) it is still possible to
put interesting upper bounds on these
asymmetries \cite{DIP,IMP,Noi}. 
In particular, within the Standard Model we can 
exclude the possibility to observe such effects at KLOE.

\subsection{$K \to 2 \pi \gamma$ and $K \to 3 \pi \gamma$ }
\no
In processes with one photon in the 
final state it is useful to expand the decay widths in 
terms of the photon energy $E_\gamma$:
\beq
\frac{ {\rm d}\Gamma(K\to \pi\pi(\pi)\gamma) }{ {\rm d}E_\gamma }
= {a \over E_\gamma} + b + c E_\gamma +O(E_\gamma^2)~.
\label{eqr}
\eeq
QED implies that the first two terms in the expansion (i.e. the 
bremsstrahlung contribution) can be unambiguously predicted 
in terms of the corresponding non--radiative amplitude \cite{Low}
and vanish if the transition involves only neutral particles.  
On the contrary, $O(E_\gamma)$ terms are `structure dependent' and
receive contributions from the so--called
direct--emission amplitudes. 

In the framework of CHPT   bremsstrahlung 
amplitudes are non--vanishing already  at $O(p^2)$ 
whereas direct emission ones, both of electric and magnetic type, 
are generated only at $O(p^4)$ \cite{Eckerrep}. Interestingly in
four--meson processes CHPT allow to extend the concept of bremsstrahlung 
relating also the dominant $O(E_\gamma)$ effect 
in (\ref{eqr}) to the corresponding non--radiative process 
\cite{DEIN1}\footnote{Actually the 
$O(E_\gamma)$ effect of `generalized bremsstrahlung' is dominant 
only if the corresponding non--radiative amplitude is not suppressed
\protect\cite{DEIN2}.}.

A complete $O(p^4)$ analysis has been performed both for $K\to 2 \pi \gamma$
\cite{ENP1,DI} and $K\to 3\pi\gamma$ \cite{DEIN2} decays. The first ones
turn out to be very promising to extract information on several 
$N_i$ combinations, the others are interesting to test the  
concept of `generalized bremsstrahlung'. Unfortunately present 
data on direct emission amplitudes are not very accurate, especially
in the $K_S \to \pi^+\pi^-\gamma$ case and in all
$K\to 3\pi\gamma$ channels. KLOE will certainly improve the present 
situation and with some effort might even succeed in detecting the
very rare  $K_S \to \pi^+\pi^-\pi^0\gamma$ transition:
BR$^{\rm th.}(K_S \to \pi^+\pi^-\pi^0\gamma)\vert_{E_\gamma > 10~{\rm MeV}}
\simeq 2 \times 10^{-10} $ \cite{DEIN2}.

\subsection{Two photon decays}
\no
Decays  with two photons in the final state are very interesting in the 
framework of CHPT \cite{Eckerrep}.
\begin{figure}[t]
\begin{minipage}[t]{77mm}
    \begin{center}
       \setlength{\unitlength}{1truecm}
       \begin{picture}(3.0,6.0)
       \put(-3.1,-7.68){\includegraphics{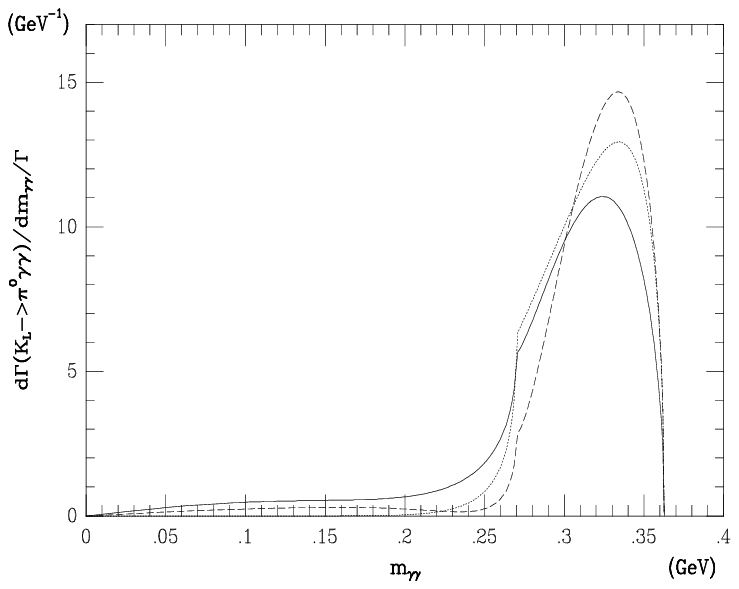}}
       \end{picture}
    \end{center}
    \caption{Theoretical predictions for the 
 width of  $K_L \rightarrow \pi^0 \gamma\gamma$
 as a function of the two--photon invariant mass. The dotted curve is the 
 $O(p^4)$ contribution,  dashed and full lines
 correspond to the  $O(p^6)$ estimates \protect\cite{Cohen,DP8} for 
 $a_V=0$ and  $a_V=-0.8$,  respectively.
 The three distributions are normalized to  
 the $50$ unambiguous events of NA31.}
\label{fig:Klpgg1}
\end{minipage}
\hspace{\fill}
\begin{minipage}[t]{77mm}
    \begin{center}
       \setlength{\unitlength}{1truecm}
       \begin{picture}(3.0,6.0)
       \put(-3.6,-3.5){\includegraphics{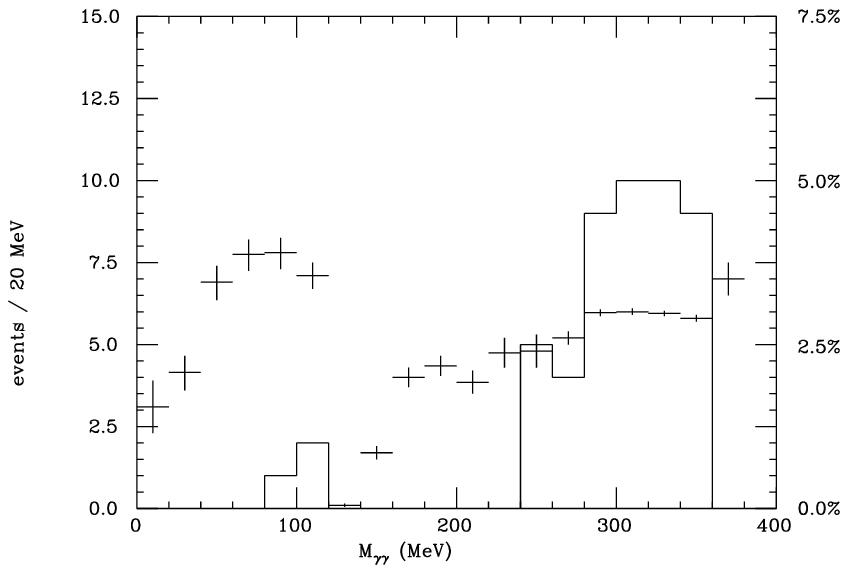}}
  \end{picture}
    \end{center}
    \caption{Distributions of the 50 unambiguously 
$K_L \rightarrow \pi^0 \gamma\gamma$ events reconstructed 
by NA31\protect\cite{NAKL} 
(histograms). Crosses indicate 
the experimental acceptance (scale on the right).}
\label{fig:Klpgg2}
\end{minipage}
\end{figure}
The first non--vanishing contribution to the decay amplitudes
of $K_S\to\gamma\gamma$ \cite{Espriu,Goity} and 
$K_L \to \pi^0 \gamma\gamma$ \cite{EPR2,Cappiello}
is of $O(p^4)$ and is generated only by (finite) loop diagrams.
The above amplitudes are thus 
unambiguously predicted in terms of $G_8$ and $G_{27}$.
In the $K_S$ case the theoretical branching ratio
is in good agreement with the experimental result
\cite{BarrKs}, on the contrary   
BR$^{\rm exp}(K_L \to \pi^0 \gamma\gamma)/$
BR$^{O(p^4)}(K_L \to \pi^0 \gamma\gamma) \gsim 2$. The reason of this 
discrepancy can be traced back to the problem of $O(p^6)$  
unitarity corrections and resonance contributions which
affect $K_L \to \pi^0 \gamma\gamma$ but not
$K_S\to\gamma\gamma$ (see e.g. Refs.~\cite{Cohen,CDM,kamborh}). 
Indeed, in $K_S\to\gamma\gamma$ 
unitarity corrections to the $K\to \pi\pi$ 
vertex are already included in the constant $G_8$ and 
vector--meson contributions are forbidden by 
kinematics. The estimate of the $O(p^6)$ local terms in  
$K_L\to \pi^0 \gamma\gamma$ is model dependent and 
a good resolution on the diphoton energy spectrum is needed
to distinguish among various models  
(Figs.~\ref{fig:Klpgg1}--\ref{fig:Klpgg2}). 

Also in $K^+\to \pi^+ \gamma\gamma$
the decay amplitude is at least of $O(p^4)$ and
the sum of the $O(p^4)$ loop diagrams is finite, however 
in this case there is an additional $O(p^4)$  contribution 
from counterterms \cite{EPR3}. Preliminary results 
form the BNL--E787 experiment  \cite{Shinkawa} 
indicate that the diphoton energy spectrum of 
this decay is consistent with the CHPT 
prediction for a reasonable value of the 
counterterm combination. 
In analogy to the $K_L \to \pi^0 \gamma\gamma$ case, it is possible 
that $O(p^6)$ effects play an important role also in 
$K^+\to \pi^+ \gamma\gamma$: recently 
D'Ambrosio and Portol\'es have shown that unitarity 
corrections are large \cite{DP6}, at least as in 
$K_L \to \pi^0\gamma\gamma$, whereas  
local contributions are likely to be suppressed \cite{DP8}.

Substantially different from the previous decays
are the $K_L \to \gamma \gamma$ and 
$K_L \to \gamma^*\gamma^*$ transitions.
Here at $O(p^4)$ there are no loop contributions and the $\pi^0$-- 
and $\eta$--pole diagrams cancel each other  
due to the Gell--Mann--Okubo relation. In general 
$O(p^6)$ effects are strongly model dependent, nevertheless 
a interesting and consistent picture of both $K_L \to \gamma \gamma^*$
and $K\to \pi\gamma\gamma$ decays has been recently proposed 
\cite{DP8}. It is worthwhile to stress that a better understanding, 
both from the theoretical and the experimental point of view, of 
$K_L \to \gamma^* \gamma^*$ form factors could help to 
disentangle interesting short--distance effects in $K_L\to\mu^+\mu^-$
\cite{Littenberg}. 

\subsection{$K\to\pi l^+l^- $ decays}
\no
The decay amplitudes of $K^+\to\pi^+ l^+l^-$ and
$K_S\to\pi^0 l^+l^-$ are at least of $O(p^4)$
and receive contributions from both loops and counterterms \cite{EPR1}.
Up to date only the two $K^+$ decay channels have been observed.
The measurement of the energy spectrum of the lepton pair 
in $K^+\to\pi^+ e^+e^-$ \cite{Alliegro} 
let to determine the $N_i$ combination
which appears also in $K^+\to\pi^+ \mu^+\mu^-$
and thus to predict unambiguously 
BR$(K^+\to\pi^+ \mu^+\mu^-)=6.2^{+0.8}_{-0.6}\times 10^{-8}$ 
\cite{EPR1,Pich2}. The preliminary measurement of 
BNL--E787 \cite{Shinkawa} is in good agreement with the 
above prediction. Unfortunately the $N_i$  combination which  
appears in $K_S\to\pi^0 e^+e^-$ is different and thus we cannot 
predict unambiguously this decay. Without model dependent
assumptions we can state only 
\cite{Eckerrep,Noi}\footnote{The counterterm combinations of 
$K_S\to\pi^0 e^+e^-$ and $K^+\to\pi^+ \mu^+\mu^-$
are correlated within specific hadronization hypotheses, 
in a wide class of models \protect\cite{EPR1,Bruno} one gets
${\rm BR}(K_S\to\pi^0 e^+e^-)\simeq 5\times 10^{-10}$.}  
\beq
10^{-10} \lsim {\rm BR}(K_S\to\pi^0 e^+e^-) \lsim 10^{-8}~.
\label{KS}
\eeq

The $K_L \to \pi^0 e^+e^-$ decay is very interesting to study
short--distance effects. Here we can distinguish 
three contributions: a short--distance dominated and
direct $CP$--violating term \cite{Buraskl}, 
the indirect  $CP$--violating 
process $K_L \to K_S\to\pi^0 e^+e^-$
and the $CP$--conserving two-photon re--scattering 
$K_L \to \pi^0 \gamma \gamma \to \pi^0 e^+e^-$.
The short--distance 
contribution can be calculated unambiguously and yields \cite{Buraskl}
\beq
{\rm BR}_{CP-dir}(K_L \to \pi^0 e^+ e^- ) =  (4.5\pm 2.6) \times 10^{-12},
\label{CPdir}
\eeq
where the error is dominated by the poor knowledge of the
CKM matrix elements involved. On the contrary the other two contributions
are affected by large theoretical
uncertainties. According to
Eq.~(\ref{KS}) the $K_S$ contribution lies between 
$3\times 10^{-13}$ and $3\times 10^{-11}$. The
absorptive part of the photon re--scattering yields
\cite{Cohen,Donoghue,DP8}
\beq
{\rm BR}_{CP-cons}(K_L \to \pi^0e^+e^-)\big|_{abs} =
(0.3 - 1.8) \times 10^{-12}, 
\eeq
where the error is related to the uncertainty on the 
diphoton spectrum of $K_L \to \pi^0 \gamma\gamma$
at small $m_{\gamma\gamma}$, the dispersive part 
--expected to be of the same order-- 
is even more model dependent \cite{Donoghue}. 

If the indirect $CP$--violating and the $CP$--conserving
contributions   were calculable with reasonable accuracy, or better if 
it was found that are negligible with respect to the
direct $CP$--violating one, an observation of 
$K_L \to \pi^0e^+e^-$
(within the reach of  future facilities \cite{Littenberg}) would
provide an important window on short--distance physics. 
To this purpose KLOE plays an important role. Indeed this 
experiment should be able both to measure BR$(K_S \to \pi^0e^+e^-)$
(or to put an upper bound on it at the level 
of $10^{-9}$) and to improve our knowledge
on the diphoton spectrum of  $K_L \to \pi^0 \gamma\gamma$.

\section{Conclusions}
\no
In this short overview we have discussed different aspects of kaon decays 
with particular attention to those cases where new measurements could 
provide important theoretical insights. The main points of our discussion
are briefly summarized in Tab.~\ref{tab:1} and we will not repeat them 
here.  Our discussion was rather qualitative and far from being complete.
Nonetheless we hope to have outlined some of the unique features of 
kaon decays in testing the Standard Model at different energy scales and, 
within this context, the important role foreseen for KLOE.

\section*{Acknowledgments}
\no
It is a pleasure to thank G. D'Ambrosio, G. Ecker and 
H. Neufeld for the fruitful collaborations on radiative kaon decays.
I am grateful to the organizers of DAPHCE96 for the invitation to 
present this talk and I acknowledge interesting discussions with 
many Conference participants. 
Last but not least, I am grateful to
J. Portol\'es for useful discussions and constructive comments on the 
manuscript.


\begin{thebibliography}{99}
\bibitem{Aloisio} A. Aloisio et al.,    
  `The KLOE Detector: Technical Proposal',
   LNF-93-002-IR, Jan. 1993.
\bibitem{Handbook} `The Second DA$\Phi$NE Physics Handbook',  
  eds. L. Maiani, G. Pancheri and N. Paver (Frascati, 1995).
\bibitem{Ecker} G. Ecker,  Prog. Part. Nucl. Phys. {\bf 35} (1995) 1. 
\bibitem{Pich} A. Pich,  Rep. Prog. Phys. {\bf  58} (1995) 563. 
\bibitem{deRafael} E. de Rafael, in 
  `CP Violation and the limits of the Standard Model', TASI 1994
   proceedings, ed. J.F. Donoghue (World Scientific, 1995).
\bibitem{Noi} G. D'Ambrosio and G. Isidori, 
  `CP violation in kaon decays', INFNNA-IV-96-29, LNF-96/033(P),
hep-ph/9611284, to appear in Int. J. Mod. Phys. A.
\bibitem{Leutwyler} H. Leutwyler, these proceedings.
\bibitem{Pennington} M. Pennington, these proceedings.
\bibitem{Maiani} L. Maiani, these proceedings.
\bibitem{GL} J. Gasser and H. Leutwyler,  \NP{B250}{85}{465}.
\bibitem{semi} J. Bijnens, G. Ecker and J. Gasser, 
  in \protect\cite{Handbook}. 
\bibitem{semi2} J. Bijnens, G. Colangelo, G. Ecker and J. Gasser, 
  in \protect\cite{Handbook}.
\bibitem{Gasserl0} J. Gasser, lectures given at the `1996 LNF Spring School
in Nuclear and Subnuclear Physics', LNF-96/030 (IR).
\bibitem{Bijnens} J. Bijnens, G. Colangelo, G. Ecker, J. Gasser 
  and M.E. Sainio,  \PL{B374}{96}{210}.
\bibitem{Stern} M. Knecht, B. Moussallam, J. Stern and N.H. Fuchs, 
  \NP{B457}{95}{513}.
\bibitem{PDG} Particle Data Group, \PR{D54}{96}{1}.
\bibitem{Ross} L. Rosselet et al., \PR{D15}{77}{574}.
\bibitem{Buras} A.J. Buras, M. Jamin, M.E. Lautenbacher and P.H. Weisz,
  Nucl. Phys. {\bf B400} (1993) 37; 
  A.J. Buras, M. Jamin and M.E. Lautenbacher, 
  ibid. {\bf B400} (1993) 75.
\bibitem{Ciuchini} M. Ciuchini, E. Franco, G. Martinelli and L. Reina, 
  Nucl. Phys. {\bf B415} (1994) 403. 
\bibitem{Fabbrichesi} M. Fabbrichesi and E.I. Lashin, \PL{B387}{96}{609}.
\bibitem{EPR1} G. Ecker, A. Pich and E. de Rafael, Nucl. Phys. {\bf B291} 
 (1987) 692.
\bibitem{EPR3} G. Ecker, A. Pich and E. de Rafael, Nucl. Phys. {\bf B303} 
 (1988) 665.
\bibitem{KMWn} J. Kambor, J. Missimer and D. Wyler, Nucl. Phys. 
  {\bf B346} (1990) 17.
\bibitem{EKW} G. Ecker, J. Kambor and D. Wyler, Nucl. Phys. {\bf B394}
  (1993) 101.
\bibitem{Nellohand} L. Maiani and N. Paver, 
  in  \protect\cite{Handbook}. 
\bibitem{Eckerrep} G. D'Ambrosio, G. Ecker, G. Isidori and H. Neufeld,
  in \protect\cite{Handbook}. 
\bibitem{Cronin}   J.A. Cronin,  \PR{161}{67}{1483}.
\bibitem{KMWlett} J. Kambor, J. Missimer and D. Wyler, Phys. Lett. 
 {\bf B261} (1991) 496.
\bibitem{DIPP} G. D'Ambrosio, G. Isidori, N. Paver and A. Pugliese, 
 Phys. Rev. {\bf D50} (1994) 5767, erratum, ibid. {\bf 51} (1995) 3975.
%
\bibitem{Kamborslopes} J. Kambor, J.F. Donoghue, B.R. Holstein, J. Missimer 
  and D. Wyler, \PRL{68}{92}{1818}.
\bibitem{DamPav} G. D'Ambrosio and N. Paver, \PR{D46}{92}{352};
  ibid. {\bf D49} (1994) 4560.
\bibitem{E621_cons} G.B. Thomson et al., \PL{B337}{94}{411}.
\bibitem{CPLEAR_cons} R. Adler et al. (CPLEAR 
  Collaboration),  \PL{B374}{96}{313}. 
\bibitem{DIP} G. D'Ambrosio, G. Isidori and N. Paver, Phys. Lett. {\bf B273}
 (1991) 497. 
\bibitem{IMP} G. Isidori, L. Maiani and A. Pugliese, Nucl. Phys. {\bf B381} 
 (1992) 522. 
%
\bibitem{Low} F.E. Low, \PR{110}{58}{974}. 
\bibitem{DEIN1} G. D'Ambrosio, G. Ecker, G. Isidori  and H. Neufeld, 
\PL{B380}{96}{165}.
\bibitem{ENP1} G. Ecker, H. Neufeld and A. Pich, Phys. Lett. {\bf B278} (1992)
 337; Nucl. Phys. {\bf B413} (1994) 321. 
\bibitem{DI} G. D'Ambrosio and G. Isidori, Z. Phys. {\bf C65} (1995) 649. 
\bibitem{DEIN2} G. D'Ambrosio, G. Ecker, G. Isidori  and H. Neufeld, 
`$K\to \pi\pi\pi\gamma$ in Chiral Perturbation Theory',
INFNNA-IV-96/47, UWthPh-1996-55, LNF-96/070(P), hep-ph/9612412.
%
\bibitem{Espriu} G. D'Ambrosio and D. Espriu, Phys. Lett. {\bf B175}
 (1986) 237. 
\bibitem{Goity} J.L. Goity, \ZP{C34}{87}{341}.
\bibitem{BarrKs} G.D. Barr et al. (CERN-NA31), \PL{B351}{95}{579}. 
%
\bibitem{EPR2} G. Ecker, A. Pich and E. de Rafael, Phys. Lett. {\bf B189}
 (1987) 363. 
\bibitem{Cappiello} L. Cappiello and G. D'Ambrosio, Nuovo Cimento
 {\bf 99A} (1988) 155. 
\bibitem{Cohen} A.G. Cohen, G. Ecker and  A. Pich,  \PL{B304}{93}{347}.
\bibitem{CDM} L. Cappiello, G. D'Ambrosio and M. Miragliuolo, 
 Phys. Lett. {\bf B298} (1993) 423.
\bibitem{kamborh} J. Kambor and  B. R. Holstein, \PR{D49}{94}{2346}. 
\bibitem{DP8} G. D'Ambrosio and J. Portol\'es,  
   `Vector meson exchange contributions to $K \to \pi \gamma \gamma$ and
   $K_L \to \gamma l^+ l^-$', INFNNA-IV-96/21, hep-ph/9610244.
\bibitem{NAKL} G.D. Barr et al. (CERN-NA31), \PL{B284}{92}{440}.
%
\bibitem{Shinkawa} T. Shinkawa (BNL--787),  `$K^+\to \pi^+ \nu\bar{\nu}$
search', talk presented at the Workshop on K Physics 
(Orsay, May 30 -- June 4, 1996). 
\bibitem{DP6} G. D'Ambrosio and J. Portol\'es, \PL{B386}{96}{403},
erratum,  ibid. {\bf 389} (1996) 770.
\bibitem{Littenberg}  L. Littenberg, `Rare kaon decay experiments,
summary and prospects', talk presented at the Workshop on K Physics
(Orsay, May 30 -- June 4, 1996).
%
\bibitem{Alliegro} C. Alliegro et al., \PRL{68}{92}{278}.
\bibitem{Bruno}  C. Bruno and J. Prades, \ZP{C57}{93}{585}.
\bibitem{Buraskl} A. Buras, M.E. Lautenbacher, M. Misiak and 
  M. Munz, \NP{B423}{94}{349}.
\bibitem{Pich2} A. Pich, `Rare kaon decays', 
talk presented at the Workshop on K Physics
(Orsay, May 30 -- June 4, 1996), hep-ph/9610243.
\bibitem{Donoghue} J.F. Donoghue and F. Gabbiani, \PR{D51}{95}{2187}.
%
\end{thebibliography}
\end{document}